\documentclass[prc,aps,showpacs,floatfix,nofootinbib,letterpaper,reprint]{revtex4-1}
\usepackage{epsfig} 
\usepackage{amssymb} 
\usepackage{amsmath} 
\usepackage{amsfonts} 
\usepackage{color, graphicx} 
\usepackage{bm}% bold math 

\def\be{\begin{equation}} 
\def\ee{\end{equation}} 

\def\Tr{{\rm Tr}}

\newcommand{\pd}[2]{\frac{\partial #1}{\partial #2}} 
\newcommand{\pdd}[2]{\frac{\partial^{2} #1}{\partial #2^{2}}} 
\newcommand{\ket}[1]{\left | #1 \right >}

\begin{document}

\title{State densities of heavy nuclei in the static-path plus random-phase approximation}
\author{P. Fanto and Y. Alhassid}
\affiliation{Center for Theoretical Physics, Sloane Physics Laboratory, Yale University, New Haven, Connecticut 06520, USA}
\date{\today}

\begin{abstract}
Nuclear state densities are important inputs to
statistical models of compound-nucleus reactions.  
State densities are often calculated with self-consistent mean-field approximations that do not include important correlations and have to be augmented with empirical collective enhancement factors.
Here, we benchmark the static-path plus random-phase approximation (SPA+RPA) to the state density in a chain of samarium isotopes $^{148-155}$Sm against exact results (up to statistical errors) obtained with the shell model Monte Carlo (SMMC) method.  
The SPA+RPA method  incorporates all static fluctuations beyond the mean field together with small-amplitude quantal fluctuations around each static fluctuation.
Using a pairing plus quadrupole interaction, we show that the SPA+RPA state densities agree well with the exact SMMC densities for both the even- and odd-mass isotopes. 
For the even-mass isotopes, we also compare our results with mean-field state densities calculated with the finite-temperature Hartree-Fock-Bogoliubov (HFB) approximation.
 We find that the SPA+RPA repairs the deficiencies of the mean-field approximation associated with broken rotational symmetry in deformed nuclei and the violation of particle-number conservation in the pairing condensate.  
In particular, in deformed nuclei  the SPA+RPA reproduces the rotational enhancement of the state density relative to the mean-field state density.
\end{abstract}


\maketitle

\section{Introduction}\label{introduction}

Nuclear level densities, which measure the average number of nuclear levels per unit energy,  are important inputs to the statistical Hauser-Feshbach theory of compound-nucleus reactions~\cite{hauser1965,koning2012}.  Neutron capture rates, which affect the predicted isotopic abundances in $r$-process nucleosynthesis~\cite{mumpower2016, surman2014} and the precision of $i$-process simulations~\cite{denissenkov2018}, are particularly sensitive to level densities.
 Uncertainties in nuclear reaction rates also have important implications for nuclear technology and stockpile stewardship applications~\cite{carlson2017}.

Level densities are extracted from various experimental data, including level counting at low excitation energies, neutron resonance data at the neutron separation energy, the Oslo method~\cite{schiller2000,voinov2001}, and particle evaporation spectra~\cite{voinov2019}.  
Rare-isotope beam facilities and novel experimental techniques such as the 
$\beta$-Oslo method~\cite{spyrou2014} promise to extend level density measurements to unstable nuclei.
However, available data to determine nuclear level densities remain limited, and theoretical calculations of level densities are required to describe many compound-nucleus reactions.% 

The calculation of level densities in the presence of correlations is a challenging many-body problem.
Most theoretical approaches are based on phenomenological models 
fitted to experimental data~\cite{koning2012,herman2007}.  
Such models cannot  be reliably extrapolated to regions in which data are scarce.
Moreover, these models are limited by uncertainties in the experimental data~\cite{voinov2019}.

Consequently, it is important to develop microscopic models of the level density that are based on the underlying nuclear interaction.
Widely used methods that rely on mean-field approximations~\cite{hilaire2006,goriely2008,hilaire2012}
must be augmented with phenomenological models to describe rotational and vibrational enhancements~\cite{goriely2008,hilaire2012}.  
In contrast, the configuration-interaction (CI) shell model 
describes both single-particle and collective excitations within the same framework.
However, conventional CI shell model methods are limited by the combinatorial growth of the many-particle model space with the number of nucleons and/or the number of single-particle states. The shell-model Monte Carlo (SMMC) method is capable of calculating level densities exactly (up to controllable statistical errors) in model spaces that are far beyond the reach of conventional CI methods, and has been applied to nuclei as heavy as the lanthanides~\cite{alhassid2008,ozen2013,bonett2013,ozen2015}; for a review, see Ref.~\cite{alhassid_rev}.
Most applications of the SMMC to the calculation of level densities have used effective nuclear interactions with a good Monte Carlo sign~\cite{alhassid_rev, lang1993,alhassid1994}, although smaller bad-sign components of more general effective nuclear interactions can be treated using the method of Ref.~\cite{alhassid1994}.  

Another method to calculate level densities based on the CI shell mode is the moment method~\cite{Mon1975,Horoi2004,senkov2016}.  This approach has been applied to light and mid-mass nuclei but becomes costly in heavy nuclei and is limited by the need to calculate independently an accurate ground-state energy.  Similarly, the stochastic estimation method of Ref.~\cite{shimizu2016} and the methods of Ref.~\cite{ormand2020} based on the Lanczos algorithm have been used to calculate CI shell model level densities in mid-mass nuclei, but these methods cannot be applied to heavy nuclei due to their prohibitive computational cost. 

The static-path plus random-phase approximation (SPA+RPA) is a promising method for including correlation effects beyond the mean field within the CI shell model framework. % that 
The approach includes large-amplitude static fluctuations beyond the mean field and small-amplitude time-dependent quantal fluctuations around each static fluctuation~\cite{kerman1981, lauritzen1991, puddu1990,lauritzen1990,puddu1991,puddu1993,attias1997,rossignoli1998}.  
In solvable models, the SPA+RPA was found to give nearly exact results for thermodynamic quantities above a low temperature below which the method breaks down \cite{puddu1990, lauritzen1990, puddu1991, attias1997, rossignoli1998}.
The SPA+RPA can also be applied to certain interactions for which the SMMC has a sign problem and thus can be complementary to the SMMC.
However, there have been few applications of the SPA+RPA to many-particle systems with realistic forces.  
In nuclear physics, the SPA+RPA was used in Ref.~\cite{rossignoli1998} to calculate thermal quantities in erbium isotopes with a pure pairing force.  The SPA+RPA was used to study the pairing properties of molybdenum isotopes with a pairing force~\cite{kaneko2006}, as well as the level density of $^{56}$Fe with a pairing plus quadrupole interaction~\cite{kaneko2007}.  The SPA+RPA has also been applied to calculate thermodynamic properties, such as the heat capacity and spin susceptibility, in nanoscale metallic grains in the presence of pairing and exchange correlations~\cite{nesterov2013}.   However, the SPA+RPA has not been benchmarked systematically against exact results in heavy nuclei.  

Here, we benchmark SPA+RPA nuclear state densities\footnote{In the state density, all $2J+1$ degenerate states associated with a nuclear level of spin $J$ are counted, whereas in the level density each level with spin $J$ is counted only once.}
 against SMMC state densities for a chain of samarium isotopes $^{148-155}$Sm, which describe the crossover from vibrational to rotational collectivity~\cite{ozen2013,gilbreth2018,mustonen2018}.   
For the even-mass samarium isotopes, we also compare our results with mean-field state densities calculated with the finite-temperature Hartree-Fock-Bogoliubov (HFB) approximation. 
We implement a Monte Carlo method to calculate SPA+RPA thermodynamic observables.  The SPA+RPA method breaks down below a certain low temperature, and we use the partition function extrapolation method of Ref.~\cite{ozen2020} to extract the ground-state energy from the SPA+RPA excitation partition function above the breakdown temperature. 

Using a pairing plus quadrupole interaction,
we find that the SPA+RPA canonical entropy and state density are in good agreement with the corresponding SMMC results for each of the even- and odd-mass samarium isotopes.
In the even-mass  deformed isotopes,  the SPA+RPA method reproduces the enhancement of the state density relative to the HFB density due to rotational collectivity.  
In addition, the SPA+RPA entropy remains nonnegative at low temperatures, whereas the pairing phase of the HFB approximation is characterized by an unphysical negative entropy.
We study the evolution with neutron number of the enhancement of the SPA+RPA and SMMC state densities relative to the HFB density, as was done in Ref.~\cite{ozen2013}.  
We find that the SPA+RPA enhancement factors are in good agreement with those extracted from the SMMC and are consistent with a crossover from pairing-dominated collectivity to rotational collectivity.  

The outline of this paper as follows.  
In Sec.~\ref{spa_rpa}, we derive the SPA+RPA expressions for the grand-canonical and approximate canonical partition functions, and we discuss the calculation of the state density.  In Sec.~\ref{practical_method}, we present the Monte Carlo method used to evaluate thermodynamic quantities in the SPA+RPA. We also summarize the partition function extrapolation method used to extract the SPA+RPA ground-state energy. In Sec.~\ref{sm_isotopes}, we apply the SPA+RPA to calculate the state densities in a chain of samarium isotopes $^{148-155}$Sm and compare our results with   
 the SMMC state densities. For the even-mass samarium isotopes, we also compare the SPA+RPA densities with the HFB densities and extract the collective enhancement factors.
Finally, in Sec.~\ref{conclusion}, we summarize our conclusions, discuss the advantages and limitations of the SPA+RPA method, and provide an outlook for future developments of this method.

\section{Static-path plus random-phase approximation to state densities}\label{spa_rpa}

\subsection{General formulation}

Here, we briefly review the SPA+RPA formalism for the partition function in the grand-canonical ensemble.
For similar derivations, see Refs.~\cite{attias1997, rossignoli1998}.  We consider a Hamiltonian in which the two-body residual interaction is written as a sum of separable terms,
\be\label{Hsep}
\hat H = \hat H_1 - \frac{1}{2}\sum_\alpha v_\alpha \hat O_\alpha^2\,,
\ee
where $\hat H_1$ is a one-body operator and $\hat O_\alpha$ are Hermitian and bilinear in fermion creation and annihilation operators.  
Any Hermitian two-body interaction can be decomposed in this way~\cite{lang1993}.  
We assume that the interaction is purely attractive when written in the form of Eq.~(\ref{Hsep}), i.e., all the $v_\alpha$ are positive.
An approximate SPA+RPA treatment of repulsive interactions was proposed in Ref.~\cite{canosa1997} but is not investigated here.

The Hubbard-Stratonovich transformation~\cite{hubbard1959,stratonovich1957} expresses the Gibbs density operator 
at inverse temperature $\beta$ as a functional integral
\be\label{hs}
e^{-\beta \hat H} = \int \mathcal{D}[\sigma(\tau)] e^{-\int_0^\beta d\tau  \sum_\alpha v_\alpha \sigma_\alpha^2(\tau)/2} \,\mathcal{T}e^{-\int_0^\beta d\tau \hat h_\sigma(\tau)}\,,
\ee
where $\sigma_\alpha(\tau)$ (with $0 \leq \tau\leq \beta$) are real-valued auxiliary fields,  
${\mathcal T}$ denotes time ordering, and $\hat h_\sigma(\tau)$ is a Hermitian one-body Hamiltonian given by
\be\label{hsigma}
\hat h_\sigma(\tau) = \hat H_1 - \sum_\alpha v_\alpha \sigma_\alpha(\tau) \hat O_\alpha\;.
\ee
Next, each auxiliary field is separated into a static and $\tau$-dependent part
\be\label{aux_sep}
\sigma_\alpha(\tau) = \sigma_\alpha + \sum_{r\neq 0} \eta_{\alpha r} e^{i\omega_r\tau}\,,
\ee
where $\omega_r = 2\pi r/\beta$ ($r=\pm 1, \pm 2,\ldots$) are the bosonic Matsubara frequencies.  The grand-canonical partition function at inverse temperature $\beta$ and chemical potentials $\mu_p,\mu_n$ (for protons and neutrons) can then be written as
\be\label{partition_fluct}
\begin{split}
Z(\beta,\mu_p,\mu_n) & = \Tr e^{-\beta (\hat H - \sum_{\lambda=p,n}\mu_\lambda \hat N_\lambda)} \\
& =   \int \left[\prod_\alpha \left(\frac{\beta v_\alpha}{2\pi}\right)^{1/2} d\sigma_\alpha \right]e^{-\beta \sum_\alpha v_\alpha \sigma_\alpha^2/2} Z(\sigma) \\
& \times \int \mathcal{D}\eta \,e^{-\beta \sum_{\alpha,r\neq0} v_\alpha |\eta_{\alpha r}|^2/2} \left \langle\mathcal{T}e^{-\int_0^\beta d\tau \hat V_\eta(\tau)}\right\rangle_\sigma\,,
\end{split}
\ee
where $Z(\sigma)$ is the partition function for static auxiliary fields $\sigma$
\be\label{Z_sp}
Z(\sigma) = \Tr e^{-\beta (\hat h_\sigma - \sum_{\lambda=p,n} \mu_\lambda \hat N_\lambda)} \,.
\ee
In Eq.~(\ref{Z_sp}), the one-body Hamiltonian $\hat h_\sigma$ is given by Eq.~(\ref{hsigma}), with the time-dependent auxiliary fields $\sigma_\alpha(\tau)$ replaced by the static fields $\sigma_\alpha$.  The operator $\hat V_\eta(\tau) = -\sum_{\alpha,r\neq0} v_\alpha \eta_{\alpha r} e^{i\omega_r\tau} \hat O_\alpha (\tau)$  in Eq.~(\ref{partition_fluct}) ($\hat O_\alpha(\tau)$ is the interaction picture representation of $\hat O_\alpha$ with respect to the static Hamiltonian $\hat h_\sigma$) accounts for the contribution of the time-dependent fluctuations of the auxiliary fields  to the one-body Hamiltonian. The angular brackets $\langle ... \rangle_\sigma$ denote the expectation value with respect to the %static grand-canonical 
static density operator $e^{-\beta (\hat h_\sigma - \sum_{\lambda=p,n} \mu_\lambda \hat N_\lambda)}$.

In the SPA+RPA, the logarithm of the integrand in Eq.~(\ref{partition_fluct}) is expanded to second order in the amplitudes $\eta_{\alpha r}$ of the time-dependent auxiliary-field fluctuations, and the resulting Gaussian integral over  $\eta_{\alpha r}$ is evaluated analytically~\cite{puddu1990,lauritzen1990,puddu1991,puddu1993, attias1997,rossignoli1998}.  The final result is 
\be\label{spa_rpa_part}
Z(\beta,\mu_p,\mu_n) = \int \left[\prod_\alpha \left(\frac{\beta v_\alpha}{2\pi}\right)^{1/2} d\sigma_\alpha \right] e^{-\beta v\cdot \sigma^2/2} Z(\sigma) C(\sigma)\,,
\ee
where $v\cdot \sigma^2 = \sum_\alpha v_\alpha \sigma_\alpha^2$ and $C(\sigma)$ is the RPA correction factor for static fields $\sigma$ given by~\cite{kerman1981, attias1997, rossignoli1998}
\be\label{rpacorr}
C(\sigma) =\frac{\prod_{k > l} \frac{1}{\tilde E_k - \tilde E_l} \sinh\left(\beta(\tilde E_k - \tilde E_l)/2\right)}{ \prod_{\nu > 0} \frac{1}{\Omega_\nu} \sinh\left(\beta \Omega_\nu/2\right)}\,.
\ee
In Eq.~(\ref{rpacorr}), $\tilde E_{k}$ are the generalized quasiparticle energies obtained by diagonalizing $\hat h_\sigma$, and $\Omega_\nu$ are the eigenvalues of the $\sigma$-dependent matrix
\be\label{rpamat}
\mathcal{S}_{kl,k^\prime l^\prime} = (\tilde E_k - \tilde E_l)\delta_{kk^\prime} \delta_{ll^\prime} - \frac{1}{2}(\tilde f_l - \tilde f_k)\sum_{\alpha} \mathcal{O}_{\alpha,kl} \mathcal{O}_{\alpha,l^\prime k^\prime}\;.
\ee
Here $\mathcal{O}_\alpha$ is the matrix representation of the one-body operator $\hat O_\alpha$ in the quasiparticle basis diagonalizing $\hat h_\sigma$, and $\tilde f_k$ is the generalized thermal quasiparticle occupation number associated with generalized quasiparticle energy $\tilde E_k$.\footnote{There are $2 N_s$ generalized quasiparticle energies and thermal occupation numbers, where $N_s$ is the number of single-particle states.  Let $k = 1,...,N_s$ and $E_k$ be the positive quasiparticle energy.  Then, $\tilde E_k = E_k$ and $\tilde E_{k+N_s} = -E_k$.  Similarly, $\tilde f_k = (1+e^{\beta E_k})^{-1}$, and $\tilde f_{k+N_s} = 1 - \tilde f_k = (1+e^{-\beta E_k})^{-1}$.}  
The matrix $\mathcal{S}$ in Eq.~(\ref{rpamat}) has the same form as the thermal RPA matrix derived from considering small oscillations around a self-consistent mean-field solution at finite temperature \cite{ring1984}.  Consequently, we refer to the eigenvalues $\Omega_\nu$ as the RPA frequencies.  These frequencies come in pairs of opposite sign, and $\prod_{\nu > 0}$ in Eq.~(\ref{rpacorr}) denotes the product over half the frequencies with a fixed sign.

Some of the RPA frequencies can become imaginary for certain static auxiliary field configurations.  The SPA+RPA is well defined at inverse temperature $\beta$ provided that there exists no imaginary frequency $\tilde \Omega_\nu$ satisfying $|\tilde \Omega_\nu| \ge 2\pi/\beta$~\cite{attias1997,nesterov2013}.  Below a certain temperature, this condition no longer holds, and the SPA+RPA breaks down.  In our calculations, we find that this breakdown temperature is very low and does not limit our ability to calculate state densities.
However, the breakdown of the SPA+RPA makes it challenging to estimate the ground-state energy, which is needed to determine the excitation energy.
We have overcome this challenge using the partition function extrapolation method~\cite{ozen2020}; see Sec.~\ref{gs}.

\subsection{Pairing plus quadrupole Hamiltonian}

Here we apply the SPA+RPA to a CI shell model Hamiltonian containing pairing and quadrupole-quadrupole interaction terms. 
The CI shell model single-particle space consists of a set of spherical orbitals $a_\lambda = (n_{a_\lambda} l_{a_\lambda} j_{a_\lambda})$, where $\lambda=p,n$ denotes protons and neutrons, respectively, and the orbitals can be different for protons and neutrons.  The orbital single-particle energies $\epsilon_{a_{\lambda}}$ correspond to a central Woods-Saxon potential plus spin-orbit interaction~\cite{BM1969}.  The Hamiltonian is given by
\be\label{Hpq}
\hat H = \sum_{a_{\lambda}} \epsilon_{a_{\lambda}} \hat n_{a_{\lambda}} - \frac{\chi}{2}\sum_{\mu=-2}^{2} :(-)^\mu \hat O_{2-\mu} \hat O_{2\mu}: - \sum_{\lambda = p,n} g_\lambda \hat P_\lambda^\dagger \hat P_\lambda\,.
\ee
In Eq.~(\ref{Hpq}), $: \; :$ denotes normal ordering and the quadrupole operator $\hat O_{2\mu} = \hat O^p_{2\mu}+ \hat O^n_{2\mu}$ is defined by $\hat O^\lambda_{2\mu} = (d{V_{\rm WS}}/{dr})_\lambda Y_{2\mu,\lambda}$, where $V_{\rm WS}$ is the Woods-Saxon central potential.  Also, $\hat P_\lambda = \sum_{(a_\lambda m) >0} c_{\,\overline{ a_\lambda m} }\, c_{a_\lambda m }$ is the pair annihilation operator, where
$\ket{a_\lambda m}$ 
is a shell-model single-particle state with magnetic quantum number $m$ for particle species $\lambda$, $(a_\lambda m ) > 0$ runs over half the states, and $\ket{\overline{ a_\lambda m }} = (-)^{j_{a_\lambda} + l_{a_\lambda} + m} \ket{a_\lambda -m}$ is the time-reversed partner of $\ket{a_\lambda m}$.  The interaction in Eq.~(\ref{Hpq}) has a similar form to the interaction used in Refs.~\cite{alhassid2008, ozen2013, bonett2013, ozen2015}, except that the latter also included octupole and hexadecupole terms in the multipole-multipole part of the Hamiltonian.

The static auxiliary fields consist of five complex fields $\alpha_{2\mu}$ ($\mu = -2,...,2$) satisfying $\alpha_{2\mu}^\ast= (-)^\mu \alpha_{2 -\mu}$ that couple to the corresponding quadrupole operators $\hat O_{2\mu}$, together with two complex fields $\Delta_{p}, \Delta_n$ that couple to the corresponding pair operators $\hat P_{p},\hat P_n$.  We transform $\alpha_{2\mu}$ to their intrinsic frame components $\tilde \alpha_{2\mu}$ defined by
\be\label{intrinsic_fields}
\begin{split}
\tilde \alpha_{20} = \beta_2\cos\gamma\,, \, \tilde\alpha_{21} = \tilde\alpha_{2-1} = 0\,,\, \tilde\alpha_{22} =\tilde \alpha_{2-2} = \frac{\beta_2}{\sqrt{2}} \sin\gamma\,,
\end{split}
\ee
and the three Euler angles characterizing the orientation of the intrinsic frame. 
 The integrand in Eq.~(\ref{spa_rpa_part})  is independent of the Euler angles  and 
 the phases of the pairing fields.  After integrating over the Euler angles and pairing field phases, the remaining integration variables are the intrinsic deformation parameters $\beta_2>0$ and $0 \le \gamma \le \pi/3$, and the pairing fields $\Delta_{p,n} > 0$. The SPA+RPA grand-canonical partition function is then given by
\be\label{part_pq}
\begin{split}
Z(\beta,\mu_p,\mu_n) & =  \int_0^{\infty} d\beta_2 \int_0^{\pi/3} d\gamma \\
& \times \int_0^\infty \int_0^\infty   d\Delta_p d\Delta_n  M(\sigma) Z(\sigma) C(\sigma)\,,
\end{split}
\ee
where $\sigma = (\beta_2,\gamma,\Delta_p,\Delta_n)$ and the measure $M(\sigma)$ is %given by
\be\label{measure}
\begin{split}
M(\sigma) & = \frac{(\beta \chi)^{5/2}}{(2\pi)^{1/2}}\left(\frac{2\beta}{g}\right)^2 \beta_2^4 \sin(3\gamma) \Delta_p \Delta_n\\
& \times e^{-\beta \chi \beta_2^2/2 - \beta \sum_{\lambda = p,n} \Delta_\lambda^2/g_\lambda}\,.
\end{split}
\ee
$Z(\sigma)$ is given by Eq.~(\ref{Z_sp}), where the static one-body Hamiltonian $\hat h_\sigma$ is
\be\label{hsigma_pq}
\begin{split}
\hat h_\sigma & = \hat H_1 - \chi \beta_2 \left[\cos \gamma\hat O_{20}  + \frac{1}{\sqrt{2}} \sin \gamma (\hat O_{22} +\hat O_{2 -2}) \right] \\
& - \sum_{\lambda=p,n}\left[ \Delta_\lambda \left(\hat P^\dagger_\lambda + \hat P_\lambda\right) + \frac{g_\lambda}{2}\left(\frac{N_{s,\lambda}}{2} - \hat N_\lambda\right)\right] \,,
\end{split}
\ee
where $\hat H_1$ is the one-body Hamiltonian in Eq.~(\ref{Hpq}) that includes a shift due to uncoupling the normal-ordered quadrupole-quadrupole term in Eq.~(\ref{Hpq}), and $N_{s,\lambda}$ is the number of single-particle states for particle species $\lambda$.  
Including the contribution of the chemical potentials and using a Bogoliubov transformation to diagonalize (\ref{hsigma_pq}) yields~\cite{nesterov2013}
\be\label{hsigma_diag}
\begin{split}
\hat h_\sigma - \sum_{\lambda=p,n} \mu_\lambda \hat N_\lambda  & = \sum_{\lambda=p,n}\sum_{k>0}\bigg[ E_{k,\lambda} (a_{k,\lambda}^\dagger a_{k,\lambda} + a_{\bar k,\lambda}^\dagger a_{\bar k,\lambda}) \\
& + (\varepsilon_{k,\lambda} - \mu_\lambda - E_{k,\lambda})\bigg]\,,
\end{split}
\ee
where $a,a^\dagger$ are annihilation and creation quasiparticle operators, $\varepsilon_{k,\lambda}$ are the eigenvalues of the particle-number-conserving term in Eq.~(\ref{hsigma_pq}), and the quasiparticle energies $E_{k,\lambda}$ are given by
\be\label{qp_energies}
E_{k,\lambda} = \sqrt{(\varepsilon_{k,\lambda} - \mu_\lambda - g_\lambda/2)^2 + \Delta_\lambda^2}\,.
\ee 
The static partition function $Z(\sigma)$ is then given by
\be\label{bcs_part}
Z(\sigma) = \prod_{\lambda = p,n} \prod_{k>0} e^{-\beta(\varepsilon_{k,\lambda} - \mu_\lambda)} 4 \cosh^2\left(\frac{\beta E_{k,\lambda}}{2}\right)\,.
\ee

The RPA correction factor $C(\sigma)$ in Eq.~(\ref{part_pq}) can be calculated using Eq.~(\ref{rpacorr}).  For the single-particle Hamiltonian (\ref{hsigma_pq}), the RPA matrix (\ref{rpamat}) connects only quasiparticle-state pairs with the same total parity and total $z$-signature.  
These symmetries render the RPA matrices block-diagonal and consequently make their diagonalization computationally more efficient.  
We note that the computational cost of diagonalizing the RPA matrix presents a significant challenge for extending the SPA+RPA method to larger model spaces and to more general interactions; see Sec.~\ref{conclusion}.

\subsection{Approximate canonical partition function}

Eq.~(\ref{part_pq}) gives the SPA+RPA partition function in the grand-canonical ensemble.  However, to calculate state densities of nuclei with given numbers of protons and neutrons, it is necessary to determine the canonical partition function~\cite{alhassid2016}. 
We do so approximately in the SPA+RPA by combining exact number-parity projection with a saddle-point approximation for particle-number projection~\cite{kaneko2007,nesterov2013}.

The number-parity projection operator is given by 
\be
\hat P_\eta= \prod_{\lambda=p,n} \frac{1 + \eta_\lambda\, e^{i \pi \hat N_\lambda}}{2}  \;,
\ee
where $\eta_p, \eta_n = +1(-1)$ for even(odd) numbers of protons or neutrons, respectively. In the SPA+RPA, the number-parity projected grand-canonical partition function is given by 
\be\label{npZ}
Z_\eta(\beta,\mu_p,\mu_n)= \int d\sigma M(\sigma) Z_\eta(\sigma) C_\eta(\sigma) \;,
\ee
where
\be\label{bcs_part_npp}
Z_\eta(\sigma) = \prod_{\lambda=p,n} \Tr\left(e^{-\beta (\hat h_{\sigma,\lambda} - \mu_\lambda \hat N_\lambda)}\right) \frac{1}{2}\left(1 + \eta_\lambda \langle e^{i\pi\hat N_\lambda} \rangle_\sigma\right)
\ee
is the number-parity projected partition function for a static auxiliary-field configuration $\sigma$, and $\hat h_{\sigma,\lambda}$ is the one-body Hamiltonian (\ref{hsigma_diag}) for particle species $\lambda$.\footnote{We note that $\hat h_\sigma$ in Eqs.~(\ref{hsigma_pq}) and (\ref{hsigma_diag}) is the sum of proton and neutron terms.}
The operator $e^{i\pi\hat N_\lambda}$ commutes with the Hamiltonian (\ref{hsigma_diag}), so Eq.~(\ref{bcs_part_npp}) can be evaluated explicitly~\cite{alhassid2005} to give
\be\label{npp_term}
\langle e^{i\pi\hat N_\lambda} \rangle_\sigma = \prod_{k>0} \tanh^2\left(\frac{\beta E_{k,\lambda}}{2}\right) \;.
\ee
$C_\eta(\sigma)$  in Eq.~(\ref{npZ})  is the number-parity projected  RPA correction factor, which is obtained by replacing the quasiparticle occupation numbers  in the RPA matrix (\ref{rpamat}) with the number-parity-projected occupation numbers
\be\label{npp_occs}
f_{k,\lambda}^\eta = \frac{f_{k,\lambda} + \eta_\lambda \langle e^{i\pi\hat N_\lambda} \rangle_\sigma f_{k,\lambda}^\pi}{1 + \eta_\lambda \langle e^{i\pi\hat N_\lambda}\rangle_\sigma}\,,
\ee
where $f_{k,\lambda} = (1+ e^{\beta E_{k,\lambda}})^{-1}$ and $f_{k,\lambda}^\pi = (1 - e^{\beta E_{k,\lambda}})^{-1}$.

The canonical partition function $Z_c(\beta,N_p,N_n)$, where $N_p$ and $N_n$ are, respectively, the number of valence protons and neutrons, is related to the number-parity-projected grand-canonical partition function $Z_\eta(\beta,\mu_p,\mu_n)$ by an inverse Laplace transform
\be\label{can_inv}
\begin{split}
Z_c(\beta,N_p,N_n) & = \frac{(2\beta)^2}{(2\pi i)^2} \int_{-i\pi/2\beta}^{i\pi/2\beta} d\mu_p d\mu_n e^{-\beta\sum_{\lambda=p,n} \mu_\lambda N_\lambda} \\ 
& \times Z_\eta(\beta,\mu_p,\mu_n) \,.
\end{split}
\ee
The leading factors of $2$ in the integrals over $\mu_p$ and $\mu_n$ in Eq.~(\ref{can_inv}) arise because the number-parity projection excludes half the possible particle numbers~\cite{rossignoli1998,rossignoli1996}.  Following Ref.~\cite{nesterov2013}, we insert Eq.~(\ref{npZ}) into Eq.~(\ref{can_inv}), change the order of the integrations over $\sigma$ and over $\mu_p,\mu_n$, and evaluate the integrals over $\mu_p,\mu_n$ by applying the saddle-point approximation to the unprojected single-particle partition function (\ref{Z_sp}). We find
\be\label{spa_rpa_part_final}
\begin{split}
Z_c(\beta,N_p,N_n) \approx \int d\sigma & M(\sigma)Z_\eta(\sigma) \\
& \times e^{ \sum_{\lambda=p,n} (\ln \zeta_\lambda - \beta\mu_\lambda N_\lambda)} C_\eta(\sigma)\,,
\end{split}
\ee
where
\be\label{zeta_def}
\zeta_\lambda = 2 \left(\frac{2\pi}{\beta^2}\pdd{\ln Z(\sigma)}{\mu_\lambda}\right)^{-1/2}\;,
\ee 
and~\cite{nesterov2013,alhassid2005}
\be\label{d2lnZdmu2}
\frac{\partial^2 \ln Z(\sigma)}{\beta^2 \partial \mu_\lambda^2} = \sum_{k>0} \frac{\beta E_{k,\lambda} (\varepsilon_{k,\lambda} - \mu_\lambda - \frac{g_\lambda}{2})^2 + \Delta_\lambda^2 \sinh(\beta E_{k,\lambda})}{2\beta E_{k,\lambda}^3 \cosh^2\left(\frac{\beta E_{k,\lambda}}{2}\right)}\;.
\ee
The chemical potential $\mu_\lambda$ in (\ref{d2lnZdmu2}) for each particle species $\lambda$ is determined by the saddle-point condition~\cite{nesterov2013}
\be\label{chm_eq}
N_\lambda = \sum_{k>0} \left[ 1- \left(\frac{\varepsilon_{k,\lambda} - \mu_\lambda - \frac{g_\lambda}{2}}{E_{k,\lambda}}\right) \tanh\left(\frac{\beta E_{k,\lambda}}{2}\right)\right]\,.
\ee

\subsection{State density}
The state density is the inverse Laplace transform of the canonical partition function
\be\label{state_density_full}
\rho(E,N_p,N_n) = \frac{1}{2\pi i}\int_{-i\infty}^{i\infty} d\beta\, e^{\beta E} Z_c(\beta,N_p,N_n) \,.
\ee
We determine the average state density by evaluating the integral in Eq.~(\ref{state_density_full})  in the saddle-point approximation~\cite{alhassid2016}
 \be\label{state_density}
\rho(E,N_p,N_n) \approx \left(\frac{2\pi}{\beta^2}C(\beta)\right)^{-1/2} e^{S_c(\beta)}\,,
\ee
where 
\be\label{entropy}
S_c(\beta) = \beta E_c(\beta) + \ln Z_c(\beta) 
\ee
is the canonical entropy and $C = dE_c/dT$ is the canonical heat capacity.\footnote{For simplicity of notation, we have omitted the explicit dependence on $N_p, N_n$ in $Z_c$, $E_c$, $S_c$, and $C$.} 
In Eq.~(\ref{state_density}), $\beta$ is a function of $E$ determined by the
saddle-point condition 
\be\label{E_saddle}
E = E_c(\beta) = -\pd{\ln Z_c(\beta)}{\beta} \;.
\ee

\section{Practical methods for calculating SPA+RPA state densities}\label{practical_method}

\subsection{Monte Carlo method}\label{mc_method}
In previous applications of the SPA+RPA, the integration over the static fields $\sigma$ was evaluated by quadrature methods~\cite{rossignoli1998, nesterov2013, kaneko2006, kaneko2007}.  
Although quadrature methods could in principle be used for the pairing plus quadrupole interaction, the computational cost of such methods scales as the exponent of the number of static fields and thus becomes prohibitive for more general effective nuclear interactions, such as the interaction used in Refs.~\cite{alhassid2008, ozen2013, bonett2013, ozen2015}.  Since our purpose is to show that the SPA+RPA can be a practical alternative to existing methods for calculating CI shell model state densities in heavy nuclei, it is important to use a computational method that can be applied to more general interactions.
Here we introduce a Monte Carlo method to calculate the SPA+RPA canonical energy and heat capacity, from which we obtain the partition function, canonical entropy, and state density.  
We evaluate the canonical energy $E_c$ defined by Eq.~(\ref{E_saddle}) using the approximate canonical partition  function $Z_c$ in Eq.~(\ref{spa_rpa_part_final}). Using dimensionless integration variables
\be\label{aux_tr}
x = \left(\sqrt{\beta\chi}\beta_2,\gamma,\sqrt{\beta/g_p}\Delta_p,\sqrt{\beta/g_n}\Delta_n \right)\,,
\ee
we rewrite Eq.~(\ref{spa_rpa_part_final}) in the form
\be\label{spa_rpa_x}
Z_c(\beta) \approx \int dx M(x) \mathcal{Z}_{\eta}(x)\,
\ee
where 
\be\label{Z_eta_def}
\mathcal{Z}_{\eta}(x) = Z_\eta(x) e^{ \sum_\lambda (\ln \zeta_\lambda - \beta\mu_\lambda N_\lambda)} C_\eta(x)\,,
\ee
and $M(x)$ is the measure in Eq.~(\ref{measure}). $M(x)$ becomes independent of $\beta$ when expressed as a function of $x$.
Using Eq.~(\ref{spa_rpa_x}) in Eq.~(\ref{E_saddle}), we find\footnote{We note that the canonical energy $E_c(\beta) = -\partial \ln Z_c/\partial \beta$ calculated in Eq.~(\ref{E_est}) is not the same as the thermal expectation value of the Hamiltonian in the SPA+RPA.  In our approach, the basic quantity is the partition function, and the canonical energy is defined by the saddle-point condition (\ref{E_saddle}).}
\be\label{E_est}
E_c(\beta)  =- \frac{\int dx W(x)\left[C_\eta(x) \pd{ \ln \mathcal{Z}_\eta(x)}{\beta}\right]}{\int dx W(x) C_\eta(x)}\,,
\ee
where $W(x)$ is the positive-definite weight function 
\be\label{weight_function}
W(x) =  \frac{M(x)\mathcal{Z}_\eta(x)}{C_\eta(x)} = M(x)  Z_\eta(x) e^{ \sum_\lambda (\ln \zeta_\lambda - \beta\mu_\lambda N_\lambda)} \,.
\ee
Eq.~(\ref{E_est}) can be rewritten as
\be\label{E_weight}
E_c(\beta) =  -\frac{\left\langle C_\eta(x)\pd{ \ln \mathcal{Z}_\eta(x)}{\beta} \right\rangle_W}{\left\langle C_\eta(x) \right\rangle_W}\,,
\ee
where the expectation value $\langle f(x)\rangle_W$ of a function $f(x)$ with respect to a weight function $W(x)$  is defined by
\be\label{W_average}
\langle f(x)\rangle_W = {\int dx W(x) f(x)\over \int dx W(x) } \;.
\ee
To evaluate the expectation values in Eq.~(\ref{E_weight}), we use a standard Monte Carlo method in which the values of $x$ are sampled according to the weight function $W(x)$.
Specifically, we perform a random walk in the space of integration variables $x$, 
updating configurations of $x$ according to the Metropolis-Hastings algorithm~\cite{gubernatis_book}.
We calculate the observables at sample configurations separated by a sufficient number of steps to ensure that the samples are decorrelated.
We then use the jackknife method \cite{young_book} to calculate expectation values and statistical errors of the thermodynamic observables, such as $E_c$ in Eq.~(\ref{E_weight}).\footnote{In practice, $\partial \ln {\cal Z}_\eta(x)/\partial \beta$ in Eq.~(\ref{E_weight}) for a given sample $x$ is evaluated using a finite-difference formula.}
Further details are provided in the Supplemental Material repository that accompanies this article~\cite{supp}.  
This Monte Carlo method is similar to the one implemented in the SMMC method~\cite{alhassid_rev}.

Having calculated $E_c(\beta)$ using Eq.~(\ref{E_weight}) for a sufficient number of $\beta$ values, we obtain the partition function by integrating Eq.~(\ref{E_saddle})
\be\label{partition_function_integral}
\ln Z_c(\beta) = S_c(0) - \int_0^\beta d\beta^\prime E_c(\beta^\prime)\,.
\ee
$S_c(0)$ in Eq.~(\ref{partition_function_integral}) is the canonical entropy at $\beta=0$ given by
\be\label{zero_beta_entropy}
S_c(0) = \sum_{\lambda = p,n} \ln \left(\begin{matrix} N_{s,\lambda} \\ N_\lambda\end{matrix}\right)\,,
\ee
where $N_{s,\lambda}$ is the dimension of the single-particle model space and $N_{\lambda}$ is the number of valence particles for particle species $\lambda$.
The heat capacity $C(\beta) = -\beta^2 \partial^2\ln Z_c/\partial \beta^2$ can also be expressed in terms of expectation values of the form (\ref{W_average}) via
\be\label{mc_hc}
\begin{split}
\pdd{\ln Z_c}{\beta} & \approx \frac{\left \langle C_\eta(x)\left[ \pdd{\ln \mathcal{Z}_\eta(x)}{\beta} + \left(\pd{\ln \mathcal{Z}_\eta(x)}{\beta}\right)^2 \right] \right \rangle_W}{\left \langle C_\eta(x)\right\rangle_W} \\
& - \left[ \frac{\left\langle C_\eta(x)\pd{ \ln \mathcal{Z}_\eta(x)}{\beta} \right\rangle_W}{\left\langle C_\eta(x) \right\rangle_W}\right]^2\,.
\end{split}
\ee
Using Eqs.~(\ref{E_weight}), (\ref{partition_function_integral}-\ref{mc_hc}) together with  Eqs.~(\ref{entropy}) and (\ref{state_density}), we calculate the canonical entropy and state density, respectively.

For importance sampling, it would have been preferable to include the RPA correction factor $C_\eta(x)$ in the weight function. However, diagonalizing the RPA matrix is the most computationally intensive part of the calculation.  We have therefore chosen $W(x)$ in Eq.~(\ref{weight_function}) that does not include $C_\eta(x)$.  
We find that the Monte Carlo method with $W(x)$ as the weight function samples the configuration space efficiently enough for our calculations.

\subsection{Ground-state energy}\label{gs}

Eq.~(\ref{state_density}) expresses the state density as a function of the absolute energy $E$.  However, in statistical reaction theory, state densities have to be known as functions of the excitation energy $E_x=E-E_0$, where $E_0$ is the ground-state energy.  However, the SPA+RPA breaks down at low but nonzero temperature, and therefore the ground-state energy cannot be calculated directly.
 A similar problem occurs in SMMC calculations in nuclei with an odd number of valence protons or neutrons with good-sign interactions, where the projection on an odd number of particles introduces a Monte Carlo sign problem at low temperatures \cite{mukherjee2012,alhassid_rev} and the ground-state energy cannot be accessed directly.  In Ref.~\cite{ozen2020}, the partition function extrapolation method was introduced to determine the ground-state energy $E_0$ of such nuclei from the calculated SMMC partition function at higher temperatures.  We apply this method, summarized below, to the SPA+RPA partition function in order to estimate the ground-state energy.

We define the excitation partition function $Z_c^\prime(\beta; E_{\rm ref})$ with respect to some reference energy $E_{\rm ref}$ by 
\be\label{Zprime_def}
Z^\prime_c(\beta; E_{\rm ref}) = Z_c(\beta)e^{\beta E_{\rm ref}}\,.
\ee
In particular, if $E_{\rm ref}=E_0$, the excitation partition function is the Laplace transform of the state density $\rho(E_x)$ given as a function of the excitation energy
\be\label{Zx_def}
Z^\prime_c(\beta; E_0) = Z_c(\beta)e^{\beta E_0} =  \int_0^\infty dE_x \rho(E_x)e^{-\beta E_x}\,.
\ee
The excitation partition function for an arbitrary reference energy is related to the excitation partition function in Eq.~(\ref{Zx_def}) by
\be\label{Zx_relation}
\ln Z^\prime_c(\beta; E_{\rm ref}) = \ln Z^\prime_c(\beta; E_0) - \beta(E_0 - E_{\rm ref})\,.
\ee

The main idea behind the partition function extrapolation method is to use a reliable model for the state density in Eq.~(\ref{Zx_relation}). For even-even nuclei, it was shown~\cite{alhassid2008} that the state density is well described by the composite formula~\cite{gilbert1965}
\be\label{composite}
\rho_{\rm comp}(E_x) = \begin{cases} e^{(E_x - E_1)/T_1} \;\;\; & E_x < E_M \\
\rho_{\rm BBF}(E_x)  & E_x > E_M
\end{cases}\,,
\ee
where $E_M$ is a matching energy, and $\rho_{\rm BBF}$ is the back-shifted Bethe formula~\cite{dilg1973}
\be\label{bbf}
\rho_{\rm BBF}(E_x) = \frac{\sqrt{\pi}}{12 a^{1/4}} \frac{e^{2\sqrt{a(E_x-\Delta)}}}{(E_x-\Delta)^{5/4}}\,.
\ee
Below $E_M$, the state density is described by the constant-temperature formula with parameters $E_1,T_1$, which are determined from the parameters  $a$ and $\Delta$ of the BBF formula by the conditions that the state density and its derivative are continuous at $E_M$.  
Inserting the composite  state density formula (\ref{composite}) into Eq.~(\ref{Zx_def}), we can fit Eq.~(\ref{Zx_relation}) to the SPA+RPA excitation partition function above the method's breakdown temperature. 

We carry out the fit in two steps \cite{ozen2020}.  In the first step, we evaluate Eq.~(\ref{Zx_def}) in the saddle-point approximation using the BBF formula for $\rho(E_x)$ to find
\be\label{func1}
\ln Z^\prime_c(\beta; E_{\rm ref}) \approx \frac{a}{\beta} + \ln\left(\frac{\pi\beta}{6a}\right) - \beta s\,,
\ee
where $s = E_0 - E_{\rm ref} + \Delta$.  Choosing $E_{\rm ref}$ sufficiently close to the yet-to-be-determined ground-state energy $E_0$, we calculate $\ln Z^\prime_c(\beta; E_{\rm ref})$ from the SPA+RPA data and fit Eq.~(\ref{func1}) to this data at moderate temperatures (for which the BBF holds).  This yields the fitted parameters $(\hat a, \hat s)$.
In the second step, we use a fit function obtained by substituting the complete composite formula $\rho_{\rm comp}(E_x)$ for $\rho(E_x)$ in Eqs.~(\ref{Zx_def}) and (\ref{Zx_relation})
\be\label{func2}
\begin{split}
\ln Z^\prime_c(\beta; E_{\rm ref}) & \approx \ln\int_0^{\infty}d E_x \rho_{\rm comp}(E_x)e^{-\beta E_x} - \beta(E_0 - E_{\rm ref})\,,
\end{split}
\ee
where $\hat a$ is fixed, $\Delta = \hat s - (E_0 - E_{\rm ref})$, and the integral is carried out numerically.  Eq.~(\ref{func2}) depends on only two parameters $E_0$ and $E_M$, which we determine with a $\chi^2$ fit.

If the backshift parameter $\Delta$ is negative, we find that the fitted value of $E_M$ is small, and the composite formula is essentially equivalent to the BBF (\ref{bbf}).  In practice, we then use $\rho_{\rm BBF}$ instead of $\rho_{\rm comp}$ in the second step of Eq.~(\ref{func2}).  We note that for negative $\Delta$,  the BBF is well defined down to $E_x=0$. 

\section{Application to samarium isotopes}\label{sm_isotopes}

Here, we apply the SPA+RPA method discussed in Secs.~\ref{spa_rpa} and \ref{practical_method} to a chain of samarium isotopes $^{148-155}$Sm, which describes the crossover from spherical to well-deformed nuclei. We use a model space consisting of the following orbitals: $0g_{7/2}$, $1d_{5/2}$, $1d_{3/2}$, $2s_{1/2}$, $0h_{11/2}$, and $1f_{7/2}$ for protons; $0h_{11/2}$, $0h_{9/2}$, $1f_{7/2}$, $1f_{5/2}$, $2p_{3/2}$, $2p_{1/2}$, $0i_{13/2}$, and $1g_{9/2}$ for neutrons.  
The selection of these orbitals is discussed in Ref.~\cite{alhassid2008}.
The single-particle energies and wave functions correspond to a Woods-Saxon central potential plus spin-orbit interaction with the parameters described in Ref.~\cite{alhassid2008}.
The quadrupole interaction parameter in Eq.~(\ref{Hpq}) is given by $\chi =  k_2\chi_0$, where $\chi_0$ is determined self-consistently~\cite{alhassid1996} and $k_2$ is a renormalization factor accounting for core polarization.  The pairing strengths $g_{p(n)} = \gamma \overline g_{p(n)}$, where $\overline g_{p(n)} = 10.9/Z(N)$ MeV ($Z$ and $N$ are the total number of protons and neutrons, respectively), and $\gamma$ is a renormalization factor.  $k_2$ and $\gamma$ are parametrized by
\be
\begin{split}
\gamma &= 1.225 \left(0.72 - \frac{0.5}{(N-90)^2 + 5.3}\right)\\
k_2 &= 2.15 + 0.0025(N-87)^2 \,,
\end{split}
\ee
This parameterization is similar to the one used in the SMMC calculations of Refs.~\cite{ozen2013,ozen2015}, except that $\gamma$ is increased by 22.5\% to account for the absence of higher-order multipoles in the interaction.

\begin{figure*}[htb!]
\includegraphics[width=\textwidth]{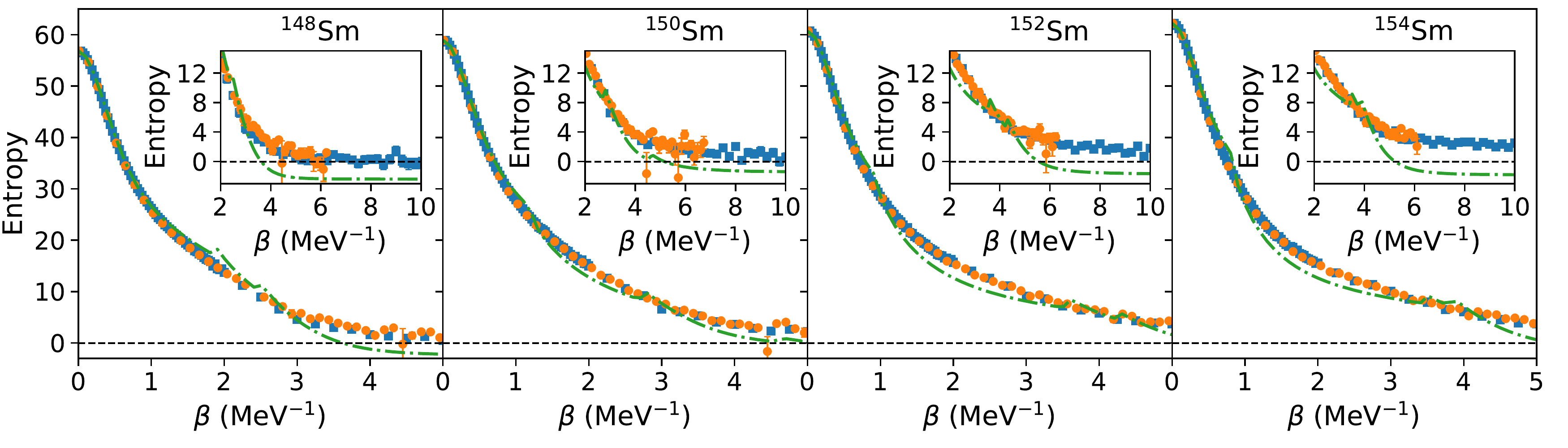}
\caption{\label{entropy_figure} The canonical entropy as a function of inverse temperature $\beta$ for the even-mass samarium isotopes $^{148,150,152,154}$Sm.  For each isotope, the SPA+RPA entropy (orange circles) is compared with the SMMC entropy (blue squares) and the HFB entropy (green dashed-dotted line).  The error bars on the SMMC and SPA+RPA results are statistical errors due to the Monte Carlo sampling. The insets show an expanded scale at large values of $\beta$.}
\end{figure*}

In the SPA+RPA, we calculate the canonical energy $E_c$  and heat capacity $C$ with the Monte Carlo method described in Sec.~\ref{mc_method}, using Eqs.~(\ref{E_weight}) and (\ref{mc_hc}), respectively. 
We define a sweep as an update of each of the four integration variables $x$.  For each $\beta$ value, we initially carry out 50 sweeps to make sure that the Monte Carlo walk is thermalized, i.e.,  the walk has reached a representative region of the configuration space according to the weight function $W(x)$.  We then calculate the observables every 70 sweeps to ensure that their values are sufficiently decorrelated.  In the Supplemental Material repository, we show that these numbers of sweeps yield acceptable thermalization and decorrelation~\cite{supp}.
We use $\sim 1000-2000$ samples per $\beta$ value in the calculation of the observables. 

From the energy $E_c(\beta)$, we calculate $\ln Z_c(\beta)$ using Eq.~(\ref{partition_function_integral}) and the canonical entropy $S_c(\beta)$  using Eq.~(\ref{entropy}). We then calculate the state density from Eq.~(\ref{state_density}).  
The Monte Carlo results and computer codes used to analyze them are provided in the Supplemental Material repository~\cite{supp}.

Below, we compare the SPA+RPA results with exact (up to statistical errors)  SMMC results and with mean-field finite-temperature HFB results.  In the SMMC, we calculate the thermal canonical energy as the expectation value of the Hamiltonian $\langle \hat H \rangle$ for fixed proton and neutron numbers; for details, see Ref.~\cite{alhassid_rev}.  We also calculate the heat capacity using the method of Ref.~\cite{liu2001}, which reduces significantly the statistical errors. Using the canonical energy and heat capacity,  we obtain the entropy and state density using Eqs.~(\ref{entropy}) and (\ref{state_density}), respectively.

For the finite-temperature HFB, we calculate the self-consistent HFB solution at each temperature using the code of Ref.~\cite{ryssens_hfshell}, together with the particle-number-projected partition function using the method of Ref.~\cite{fanto2017}.   We then apply Eqs.~(\ref{E_saddle}), (\ref{entropy}), (\ref{state_density}) to obtain, respectively, the energy, entropy, and state density.  

\subsection{Even-mass samarium isotopes}

In Fig.~\ref{entropy_figure}, we show the canonical entropy $S_c$ as a function of inverse temperature $\beta$ for the SPA+RPA (orange circles), SMMC (blue squares), and HFB (green dashed-dotted line) for the even-mass samarium isotopes $^{148,150,152,154}$Sm.  The SPA+RPA entropies are shown up to values of $\beta$ close to the value above which the approximation breaks down. For each of the isotopes, we find the SPA+RPA entropy to be in excellent agreement with the SMMC entropy.  The two kinks in the HFB entropy for $^{148}$Sm indicate the proton and neutron pairing phase transitions, and the additional kink at lower $\beta$ for the other isotopes is due to the shape phase transition from a spherical to a deformed mean-field solution.  

At $\beta$ values above the shape transition, the HFB entropy significantly underestimates the SPA+RPA and SMMC entropies because the HFB does not describe the contribution of rotational bands that are built on intrinsic mean-field band heads~\cite{alhassid2016}.  
The SPA+RPA restores the rotational symmetry that is broken in the HFB and thus reproduces this rotational enhancement of the entropy.  Furthermore, in the pairing phase, the HFB entropy becomes unphysically negative because of the inherent breaking of particle-number conservation in the HFB approximation~\cite{fanto2017}.  
In contrast, the SPA+RPA entropy remains nonnegative (within statistical errors) because the SPA+RPA repairs the intrinsic violation of particle-number conservation.  Finally, as the neutron number increases, the SPA+RPA and SMMC entropies remain nonzero to increasingly large values of $\beta$, indicating the presence of a rotational enhancement down to lower temperatures in nuclei with larger deformation. 

We used the partition function extrapolation method summarized in Sec.~\ref{gs} to estimate the ground-state energy from the SPA+RPA partition function above the breakdown temperature of the approximation.  For $^{148,150}$Sm, we used the composite formula (\ref{composite}) in the second step of the fit, whereas for $^{152,154}$Sm the back-shift parameter $\Delta$ is negative, and it was simpler to use the BBF formula (\ref{bbf}).  In Table~\ref{ground_state_table}, we compare the SPA+RPA ground-state energy estimates to the SMMC and HFB ground-state energies.  
We calculated the SMMC ground-state energies by taking a weighted average of the thermal energy at large $\beta$ values ($\beta \approx 8-20$ MeV$^{-1}$).  
Table~\ref{ground_state_table} shows that the SPA+RPA misses at most $\sim$600 keV of ground-state correlation energy, whereas the HFB misses a few MeV of correlation energy in each isotope.   The agreement between the SPA+RPA estimate and the SMMC ground-state energy improves with decreasing deformation, and the two agree with each other for the spherical isotope $^{148}$Sm.  
In Table \ref{sden_fits_table}, we show the parameters $a,\Delta, E_M$ of the state density formulas obtained from the ground-state energy fits.  

\begin{table}[h!]
\caption{\label{ground_state_table} The ground-state energies (in MeV) for the SMMC, SPA+RPA, and HFB for $^{148,150,152,154}$Sm.  The SPA+RPA estimates are obtained with the partition function extrapolation method described in Sec.~\ref{gs}, with errors arising from the $\chi^2$ fit to the SPA+RPA partition function data.  The SMMC values are obtained by taking a weighted average of the thermal energies at large $\beta$ values.}
\begin{tabular}{l c c c}
\hline\hline
% Sm148
& SMMC & SPA+RPA & HFB \\ \cline{2-4}
$^{148}$Sm \hspace{.5cm} &  -234.180 $\pm$ 0.016 & -234.131 $\pm$ 0.021 & -230.979 \\  
% Sm150
$^{150}$Sm \hspace{.5cm} & -254.019 $\pm$ 0.014 & -253.859 $\pm$ 0.015 & -251.127  \\ 
% Sm152
$^{152}$Sm \hspace{.5cm} & -273.756 $\pm$ 0.010 & -273.242 $\pm$ 0.017 & -271.153 \\ 
% Sm154
$^{154}$Sm \hspace{.5cm} & -293.292 $\pm$ 0.010 &-292.680 $\pm$ 0.017 & -290.449 \\ \hline\hline
\end{tabular}
\end{table}

\begin{table}[h!]
\caption{\label{sden_fits_table} The parameters obtained from the partition function extrapolation method discussed in Sec.~\ref{gs} applied to the SPA+RPA. 
For $^{148,150}$Sm, we used the composite formula (\ref{composite}) in the second step of the fit, while for $^{152,154}$Sm, we used the BBF (\ref{bbf}).}
\begin{tabular}{l c c c}
\hline\hline
 & $a$ (MeV$^{-1}$) & $\Delta$ (MeV) & $E_M$ (MeV)  \\ \cline{2-4}
 $^{148}$Sm \hspace{.5cm} & 17.09 $\pm$ 0.11 & 1.07 $\pm$ 0.03 & 1.452 \\
 $^{150}$Sm \hspace{.5cm}& 18.28 $\pm$ 0.06 & 0.62 $\pm$ 0.02 & 0.95 \\
 $^{152}$Sm \hspace{.5cm}& 19.14 $\pm$ 0.10 & -0.17 $\pm$ 0.03 & --\\
 $^{154}$Sm \hspace{.5cm} & 18.89 $\pm$ 0.13 & -0.38 $\pm$ 0.03 & --\\
 \hline\hline
\end{tabular}
\end{table}

\begin{figure*}[bth!]
\centering
\includegraphics[width=\textwidth]{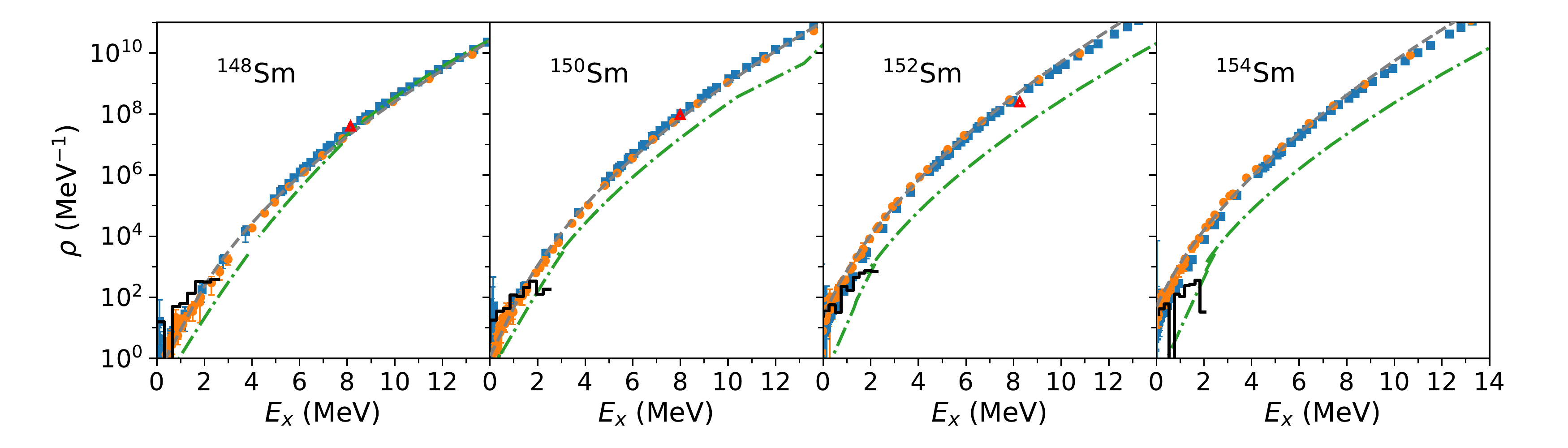}
\caption{\label{state_density_figure} The state density  $\rho$ as a function of excitation energy $E_x$ for the even-mass samarium isotopes.  The SPA+RPA state density (orange circles) is compared with the SMMC density (blue squares) and the HFB density (green dashed-dotted lines) for each isotope.  The grey dashed lines show the composite formula (\ref{composite}) for $^{148,150}$Sm and the BBF (\ref{bbf}) for $^{152,154}$Sm, with the parameters from Table~\ref{sden_fits_table} (see text).  Experimental neutron resonance data (red triangles) and low-energy level counting data (black histograms) are also shown.  The error bars show statistical errors in the SPA+RPA and SMMC results.}
\end{figure*}

Using the ground-state energies in Table \ref{ground_state_table}, we calculated the SMMC, SPA+RPA, and HFB state densities 
for $^{148,150,152,154}$Sm.  Fig.~\ref{state_density_figure} shows these densities, using a similar convention as in Fig.~\ref{entropy_figure}.  In each isotope, the SPA+RPA state density is in good agreement with the SMMC state density.   
In contrast, the HFB state density significantly underestimates the SMMC and SPA+RPA densities. 
As the neutron number increases, the enhancement of the SMMC and SPA+RPA state densities over the HFB state density persists to higher excitation energy.  
This enhancement originates in the contribution of rotational bands that are included in the SMMC and SPA+RPA densities but are not described by the HFB approximation.  We observe an additional enhancement of the SPA+RPA and SMMC state densities over the HFB density at very low excitation energies, which is due to the unphysical negative entropy in the pairing phase of the HFB. This latter enhancement is particularly apparent in the spherical nucleus $^{148}$Sm.

In Fig.~\ref{state_density_figure} we also compare the calculated state densities with experimental state densities obtained from level counting at low excitation energies~\cite{nndc} (black histograms) and the average $s$-wave neutron resonance spacings $D_0$ at the neutron threshold~\cite{ripl} (red triangles).  We used a spin cutoff model~\cite{bethe1937,ericson1960} with the rigid-body moment of inertia to convert $D_0$ values to state densities.   The agreement between the data and the SPA+RPA and SMMC state densities is good overall, in particular for $^{148,150}$Sm.  In $^{152,154}$Sm, the calculated densities overestimate the experimental data.  In contrast, the mean-field HFB densities  do not agree well with the experimental data. 

In Fig.~\ref{state_density_figure} we also show the phenomenological composite or BBF state densities calculated with the parameters reported in Table~\ref{sden_fits_table}. We find good agreement between these fitted parameterizations and the SPA+RPA state densities.  This agreement demonstrates that the partition function extrapolation method to extract the ground-state energy described in Sec.~\ref{gs} is reliable. 

\begin{figure*}[t!]
\centering
\includegraphics[width=\textwidth]{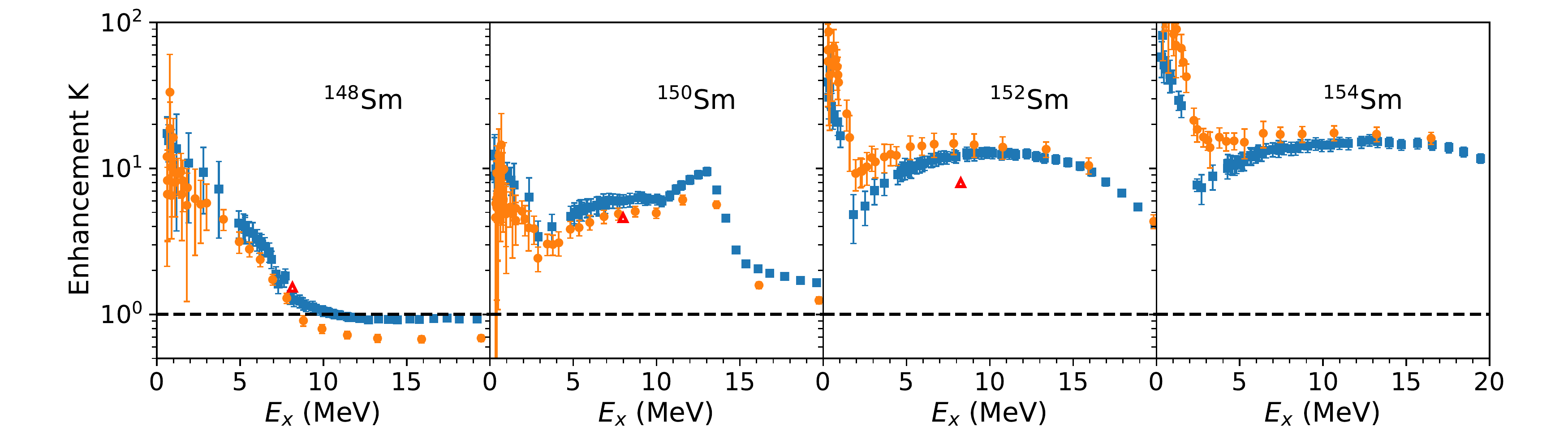}
\caption{\label{enhancement_figure} The enhancement factor $K = \rho/\rho_{\rm HFB}$ for the SPA+RPA (orange circles), the SMMC (blue squares), and the neutron resonance data (red triangles) for the even-mass samarium isotopes.  The error bars indicate statistical errors from the Monte Carlo sampling.}
\end{figure*}

\begin{figure*}[t!]
\centering
\includegraphics[width=\textwidth]{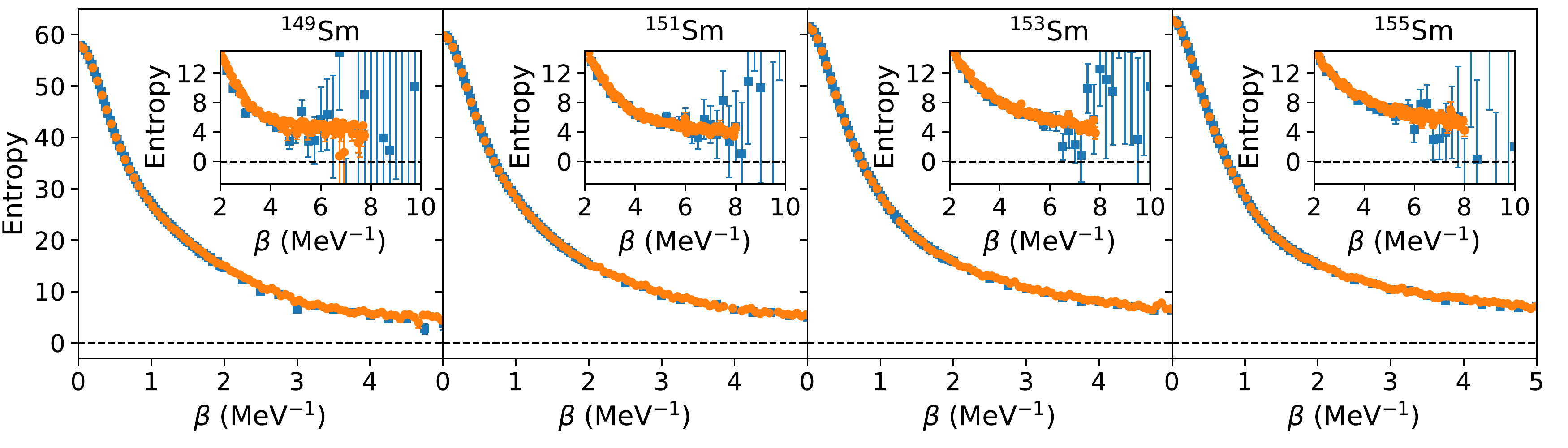}
\caption{\label{odd_entropy_figure} The canonical entropy as a function of inverse temperature $\beta$ for the odd-mass samarium isotopes $^{149,151,153,155}$Sm.  The SPA+RPA entropy (orange circles) is compared with the SMMC entropy (blue squares) for each isotope.  The error bars show statistical errors arising from the Monte Carlo sampling.  The insets show an expanded scale at large values of $\beta$.}
\end{figure*}

To demonstrate even more clearly how well the SPA+RPA describes correlations that are missing in the mean-field approximation, we show in Fig.~\ref{enhancement_figure} the state density enhancement factor $K = \rho/\rho_{\rm HFB}$ for the SMMC (blue squares) and SPA+RPA (orange circles).  The SPA+RPA enhancement factors are in good agreement with the SMMC enhancement factors.  In the spherical nucleus $^{148}$Sm, the enhancement factor differs significantly from one only at the lowest excitation energies and is due entirely to the unphysical negative entropy in the pairing phase of the HFB.  In the deformed isotopes $^{150,152,154}$Sm, a significant rotational enhancement of $\sim 10$ appears and persists to increasing excitation energy
as the neutron number increases.  This change in the enhancement factor indicates the crossover from pairing-dominated to rotational collectivity in the chain of samarium isotopes~\cite{ozen2013,gilbreth2018,mustonen2018}.  Fig.~\ref{enhancement_figure} also shows that the SPA+RPA and SMMC results are in good agreement with the neutron resonance data (red triangles) in $^{148,150}$Sm.  The calculated enhancement factors somewhat overestimate the neutron resonance data in $^{152}$Sm.

\subsection{Odd-mass samarium isotopes}

Having established the accuracy of the SPA+RPA state densities for the even-mass samarium isotopes, we next benchmark the SPA+RPA state densities for the odd-mass samarium isotopes $^{149,151,153,155}$Sm.  
In Fig.~\ref{odd_entropy_figure}, we compare the SPA+RPA canonical entropy (orange circles) with the SMMC canonical entropy (blue squares) for the odd-mass isotopes.
In each isotope, the SPA+RPA entropy is in excellent agreement with the SMMC entropy.
The odd-mass sign problem leads to large fluctuations of the SMMC entropy at high values of $\beta$, as is shown in the insets of Fig.~\ref{odd_entropy_figure}. 
The SPA+RPA entropy remains reliable to slightly lower temperatures than the SMMC entropy.  
Both the SPA+RPA and SMMC entropies appear to converge to a nonzero limit, indicating the magnetic degeneracy of the nonzero-spin ground state of the odd-mass system.

\begin{figure*}[t!]
\centering
\includegraphics[width=\textwidth]{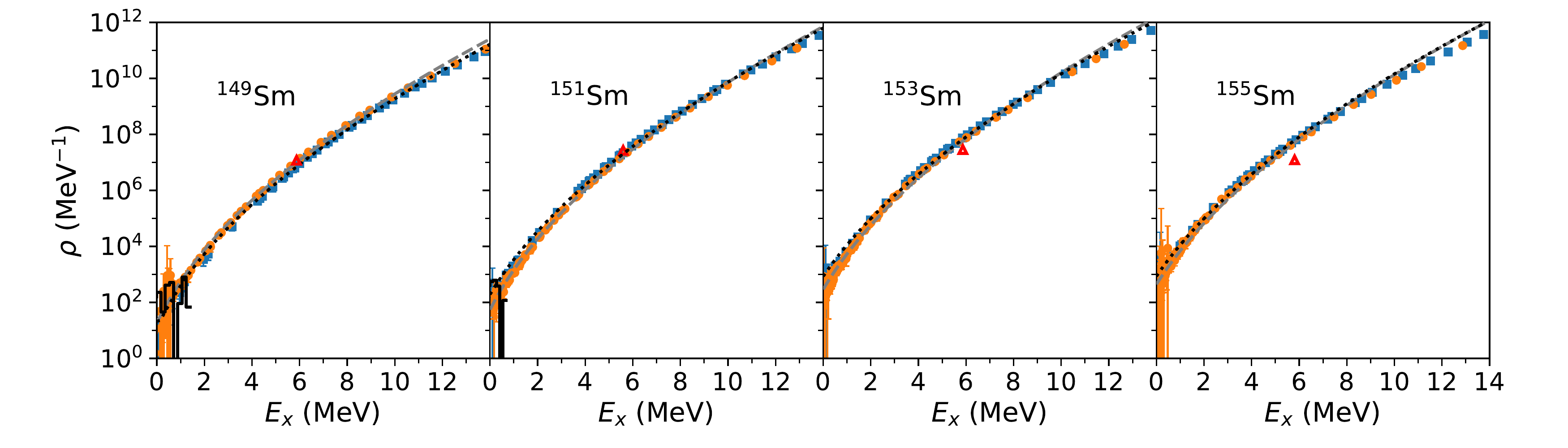}
\caption{\label{odd_state_density_figure} The state density as a function of excitation energy $E_x$ calculated with the SPA+RPA (orange circles) and SMMC (blue squares) for the odd-mass samarium isotopes $^{149,151,153,155}$Sm.  The BBF calculated with parameters obtained from fitting the SPA+RPA ground-state energies (grey dashed lines) and fitting the SMMC ground-state energies (black dotted lines) are also shown.  
Experimental neutron resonance data (red triangles) and low-energy level counting data (black histograms) are shown as well.}
\end{figure*}

The projection on the odd number of neutrons introduces a Monte Carlo sign problem in the SMMC at low temperatures~\cite{mukherjee2012,alhassid_rev} that prevents the ground-state energy from being calculated directly.  To obtain the SMMC and SPA+RPA state densities as functions of excitation energy,  we used the partition function extrapolation method, which is summarized in Sec.~\ref{gs}, to determine the ground-state energies $E_0$ in both approaches. We used the BBF (\ref{bbf}) in the second step of the fits.
 Table~\ref{odd_table} shows the extracted values of $E_0$ and the BBF state density parameters $a,\Delta$ for the SMMC and SPA+RPA.  
The agreement between the SMMC and SPA+RPA ground-state energy estimates is even better than for the even-mass isotopes, with the largest discrepancy of $\sim 300$ keV in $^{149}$Sm.

\begin{table}[h!]
\centering
\caption{\label{odd_table} Ground-state energies $E_0$ and $a,\Delta$ values from fitting the BBF (\ref{bbf}) to the SPA+RPA and SMMC excitation partition functions for the odd-mass samarium isotopes.}
\begin{tabular}{l l c c c}
\hline \hline
& & $E_0$ (MeV) & $a$ (MeV$^{-1}$) & $\Delta$ (MeV) \\ \cline{3-5}
% Sm149
$^{149}$Sm \hspace{0.0cm} & SPA+RPA & -242.957 $\pm$ 0.008 & 18.36 $\pm$ 0.04 & -0.13 $\pm$ 0.01 \\
& SMMC & -243.327 $\pm$ 0.019 & 17.97 $\pm$ 0.04 & -0.04 $\pm$ 0.02 \\ \hline\hline
%Sm151
$^{151}$Sm \hspace{0.0cm} & SPA+RPA & -262.913 $\pm$ 0.006 & 19.24 $\pm$ 0.07 & -0.39 $\pm$ 0.02 \\
& SMMC & -262.909 $\pm$ 0.047 & 18.63 $\pm$ 0.06 & -0.77 $\pm$ 0.05 \\ \hline\hline
% Sm153
$^{153}$Sm \hspace{0.0cm} & SPA+RPA & -282.384 $\pm$ 0.005 & 19.57 $\pm$ 0.12 & -0.84 $\pm$ 0.02 \\
& SMMC & -282.449 $\pm$ 0.031 & 18.78 $\pm$ .09 & -1.25 $\pm$ 0.05 \\ \hline\hline
% Sm155
$^{155}$Sm \hspace{0.0cm} & SPA+RPA & -301.949 $\pm$ 0.003 & 19.07 $\pm$ 0.12 & -1.00 $\pm$ 0.03 \\
& SMMC & -302.077 $\pm$ 0.021 & 18.27 $\pm$ 0.10 & -1.39 $\pm$ 0.04 \\ \hline\hline
\end{tabular}
\end{table}

In Fig.~\ref{odd_state_density_figure}, we compare the state densities calculated with the SMMC (blue squares) and SPA+RPA (orange circles).  The results are in good agreement with each other and with available experimental data from level counting~\cite{nndc} (black histograms) and the average $s$-wave neutron resonance spacings~\cite{ripl} (red triangles). 
The agreement between the calculated and experimental state densities degrades somewhat as the neutron number increases and is of similar quality to the agreement found in Ref.~\cite{ozen2015} using an interaction that included contributions from higher-order multipoles.
We also show in  Fig.~\ref{odd_state_density_figure} the BBF state densities calculated with the parameters tabulated in Table~\ref{odd_table}.  
These fitted BBF densities agree well with the calculated state densities. 

\section{Conclusion and outlook}\label{conclusion}

Here we benchmarked state densities calculated with the SPA+RPA in the CI shell model framework against exact (up to controllable statistical errors) SMMC state densities for a chain of samarium isotopes $^{148-155}$Sm.
 We implemented a Monte Carlo method to calculate the canonical energy and heat capacity in the SPA+RPA, from which we determined the canonical entropy and state density.
The SPA+RPA ground-state energy was estimated from the excitation partition function above the SPA+RPA breakdown temperature using the partition function extrapolation method \cite{ozen2020}.

We found good agreement between the SPA+RPA state densities and SMMC state densities for all the isotopes considered.  For the even-mass samarium isotopes,  we also calculated mean-field state densities using the finite-temperature HFB approximation. The main deficiencies of the mean-field approximation arise from the broken rotational symmetry in deformed nuclei and the inherent violation of particle-number conservation in the pairing condensate.  Consequently, the mean-field approximation cannot reproduce the contribution of rotational bands that are characteristic of deformed nuclei and yields an unphysical negative entropy in the pairing phase of the HFB. The SPA+RPA 
resolves these deficiencies of the mean-field approximation. In particular, the SPA+RPA reproduces well the rotational collective enhancement of the state density relative to the mean-field density in deformed nuclei. This enhancement persists to higher excitation energies as the neutron number increases, showing that the importance of rotational collectivity increases with deformation.  Overall, our results show that the SPA+RPA provides state densities in the CI shell model framework that are in agreement with exact SMMC densities.

A significant limitation of the SPA+RPA method is the computational cost of diagonalizing the RPA matrix at each sampled configuration of the static fields.  The dimension of the RPA matrix scales as $\sim N_{s,p}^2 + N_{s,n}^2$ (where $N_{s,p(n)}$ is the number of proton(neutron) single-particle states), and the cost of diagonalizing this matrix scales as the cubic power of this dimension.\footnote{For the case considered here, calculating the RPA correction factor $C_\eta(x)$ takes $\sim$ 3 minutes on a standard laptop (2 GHz Intel Core i5 MacBook Pro with 32 GB of RAM) and must be calculated three times per Monte Carlo sample due to the finite-difference calculation of the energy and heat capacity.}
Calculating the canonical energy and heat capacity in the SMMC scales as a lower power of the number of single-particle states, specifically as $\sim N_{s}^4$ for each particle species. 
It would therefore be useful to investigate methods for speeding up the calculation of the RPA correction factor.  
One such method was proposed in Ref.~\cite{kaneko2005}.

In comparing the SPA+RPA to the SMMC, it is also useful to consider the limits of the applicability of each method.
The SPA+RPA method requires that the single-particle Hamiltonian $\hat h_\sigma$ in Eq.~(\ref{hsigma}) be a Hermitian operator for any configuration $\sigma$ of the static auxiliary fields.
This condition is guaranteed if all terms in the Hamiltonian are attractive when written in the separable form $\hat O_\alpha^2$ of Eq.~(\ref{Hsep}), where each operator $\hat O_\alpha$ is Hermitian.
Moreover, this condition guarantees that, at temperatures above the breakdown temperature of the SPA+RPA, the weight function $W(\sigma)$ of the Monte Carlo method discussed in Sec.~\ref{mc_method} and the RPA correction factor $C_\eta(\sigma)$ are both positive definite for any static field configuration $\sigma$.
Consequently, $W(\sigma)$ can be used as a weight function to sample the static fields, and the Monte Carlo method described in Sec.~\ref{mc_method} will not have a sign problem.
In contrast, for the SMMC method to have a good Monte Carlo sign, the Hamiltonian must be invariant under time reversal, and all of its interaction terms must be attractive when written as a sum over terms of the form $\{\hat O_\alpha,  \bar O_\alpha\}$ where $\bar  O_\alpha$ is the time reverse of $\hat O_\alpha$ 
~\cite{alhassid_rev, lang1993,alhassid1994}.  It can be shown that the time-reversal and Hermitian conjugate of a tensor one-body density operator are related by a sign. Thus, in some cases, either the SPA+RPA or SMMC would have good sign while the other method would have a sign problem, and the two methods would complement each other.
Furthermore, the SPA+RPA can be applied if time-reversal symmetry is broken, e.g., in the presence 
of a cranking term $-\omega \hat J_{i}$ ($i = x,y,z$), which would cause a sign problem in the SMMC.

A method for approximately including repulsive interactions in the SPA+RPA framework was proposed in Ref.~\cite{canosa1997}.
It would be interesting to benchmark this method for realistic nuclear interactions that include repulsive components. 

Finally, statistical reaction codes require as input spin- and parity-dependent level densities, rather than just state densities.  
To calculate these level densities, it is necessary to extend the SPA+RPA formalism to include spin and parity projections. 

\section*{Acknowledgments}
We gratefully acknowedge W. Ryssens for the use of the HF-SHELL code~\cite{ryssens_hfshell} in the calculation of the HFB state densities, and for providing the parameters of the CI shell model interaction used in this work.  
We also thank G. F. Bertsch for useful discussions and for providing comments on the manuscript.
This work was supported in part by the U.S. DOE grant No.~DE-SC0019521, 
and by the U.S. DOE NNSA Stewardship Science Graduate Fellowship under cooperative agreement No.~NA-0003864.
The calculations used resources of the National Energy Research Scientific Computing Center (NERSC), a U.S. Department of Energy Office of Science User Facility operated under Contract No.~DE-AC02-05CH11231.  We thank the Yale Center for Research Computing for guidance and use of the research computing infrastructure.

\end{document}